\documentclass[lettersize,journal]{IEEEtran}
\IEEEoverridecommandlockouts
\usepackage{cite}
\usepackage{amsmath,amssymb,amsfonts}
\usepackage{algorithmic}
\usepackage{multirow}
\usepackage{graphicx}
\usepackage{textcomp}
\usepackage{xcolor}
\def\BibTeX{{\rm B\kern-.05em{\sc i\kern-.025em b}\kern-.08em
    T\kern-.1667em\lower.7ex\hbox{E}\kern-.125emX}}
\begin{document}

\title{Artifact-Robust Graph-Based Learning in Digital Pathology\\
\author{Saba Heidari Gheshlaghi, Milan Aryal, Nasim Yahyasoltani, and Masoud Ganji}
\thanks{ Saba Heidari Gheshlaghi and Milan Aryal contributed equally in this paper.\\ }
\thanks{ S. Heidari Gheshlaghi, M. Aryal and  N. Yahyasoltani are with the Department of Computer Science, Marquette University,
Milwaukee, WI 53202 USA (e-mail: \{saba.heidari, milan.aryal, nasim.yahyasoltani\}@marquette.edu). }
\thanks{M. Ganji is a pathologist with the Northshore Pathologists, S.C., Milwaukee, WI 53211 USA (e-mail: maganji@yahoo.com).}}

\maketitle

\maketitle

\begin{abstract}
Whole slide images~(WSIs) are digitized images of tissues placed in glass slides using advanced scanners. The digital processing of WSIs is challenging as they are gigapixel images and stored in multi-resolution format. A common challenge with WSIs is that perturbations/artifacts are inevitable during storing the glass slides and digitizing them. These perturbations include motion, which often arises from slide movement during placement, and changes in hue and brightness due to variations in staining chemicals and the quality of digitizing scanners. In this work, a novel robust learning approach to account for these artifacts is presented. Due to the size and resolution of WSIs and to account for neighborhood information, graph-based methods are called for.  We use graph convolutional network~(GCN) to extract features from the graph representing WSI. Through a denoiser {and pooling layer}, the effects of perturbations in WSIs are controlled and the output is followed by a transformer for the classification of different grades of prostate cancer. To compare the efficacy of the proposed approach, the model without denoiser is trained and tested with WSIs without any perturbation and then different perturbations are introduced in WSIs and passed through the network with the denoiser. The accuracy and kappa scores of the proposed model with prostate cancer dataset compared with non-robust algorithms show significant improvement in cancer diagnosis. 
\end{abstract}


\section{Introduction}
\label{sec:introduction}
Pathology glass slides are used for the diagnosis of various diseases such as cancer, infectious diseases, autoimmune disorders, and neurological conditions. The field of pathology has experienced a significant transformation with the advent of whole slide images~(WSIs), also known as digital pathology slides. WSIs are high-resolution digital representations of glass slides used in traditional pathology practice and allow pathologists and researchers to examine high-resolution tissue samples (often exceeding 100,000 pixels) \cite{ref3}. WSIs offer numerous advantages over traditional glass slides, including remote access, storage efficiency, and advanced analysis capabilities. 
However, as with any digital system, various perturbations and corruptions can compromise the quality of images. Perturbations in WSIs result in small 
alterations in the images. In addition, WSIs can be susceptible to different types of noise and adversarial perturbations, which can pose challenges for accurate diagnosis and analysis.  \\
Over the past decade, numerous studies have employed deep learning (DL) models for the purposes of diagnosis, prognosis, and classification of various diseases. Despite the high-resolution and computational complexity of the WSIs, their utilization in computational pathology and machine-based cancer diagnosis is on the rise \cite{ref4}. However, processing of high-resolution WSIs by DL models remains a challenging task. The sheer size of a WSI, with billions of pixels in a single file typically larger than a gigabyte, presents difficulties in training with common DL methods like convolutional neural network (CNN). A common approach in applying CNN models for WSIs involves dividing the image into patches and processing them separately. Nevertheless, processing patches individually may result in the loss of crucial pathological features since the surrounding features play a significant role in elucidating them.

The use of graph convolutional network (GCN) on WSIs has emerged as a promising approach to overcome the challenges associated with traditional CNN-based methods \cite{milan2,w11,ref3}. They are particularly well-suited for capturing spatial dependencies, handling irregular structures and complex relationships within the high-resolution WSIs, which are critical for accurate cancer diagnosis and classification. In \cite{milan} the authors tackle the lack of annotated data obstacle by using GCN-based self-supervised learning on WSIs. Self-supervised learning enables to extract meaningful representations from the data without relying on labeled or annotated samples. 

Studies have demonstrated that DL models, including CNNs and GCNs, are susceptible to adversarial attacks where imperceptible modifications are made to input sample.  The work in \cite{attack1} was the first work to demonstrate the vulnerability of CNNs to adversarial input samples and attacks. Since then, several studies have explored different approaches for improving DL robustness and generalization. These attacks involve adding perturbations or corruption to the input data, causing the models to make false classifications or predictions. Due to the natural presence of such perturbations, DL models can be easily fooled by corrupted or adversarial samples, leading to a significant impact on their performance and reliability. {Adversarial vulnerabilities in the context of medical images, including WSIs, are of significant concern in the field of healthcare and have wide-ranging implications for patient safety, data privacy, ethical considerations, regulatory compliance, and the overall reliability and trustworthiness of DL model in healthcare. \cite{attack1,w4,w2,adv1,adv2}.

Ensuring that DL models remain robust to changes and maintain consistent performance, even in the presence of noisy or corrupted input samples, is crucial. As regards the medical domain where accurate diagnosis, treatment planning, and patient safety are of utmost concern, robust DL models can help mitigate wrong diagnosis, and minimize the risk of false positives or negatives to ensure the integrity of patient care. A commonly employed technique to enhance model robustness is adversarial training. This method to expose models to adversarial examples during the training process, helping the models to learn more robust and discriminative features \cite{advt1,advt2}.} Other investigations have delved into incorporating advanced normalization methods or architectural modifications to mitigate the impact of adversarial perturbations. {The use of inverse imaging problems (IIPs) to reduce the effects of noise and perturbations has recently gained attention. However, these approaches require prior knowledge and pairs of noisy and denoised images, which are not consistently available. To address this challenge, the concept of untrained neural network priors (UNNPs) has emerged, enabling denosing without the need for prior information~\cite{ref8,ref9}. In UNNPs, rather than training deep learning models on extensive datasets, the model is employed to capture the essence of images. }

{There are very limited work for enhancing the robustness of cancer detection using WSIs. Those aproaches mainly involve the utilization of patch-based images and models based on CNNs.} The work in \cite{w4} demonstrated that a highly accurate model used for classifying tumor patches in pathology images can be easily fooled by corrupted samples. The study proposed a single universal perturbation matrix capable of being added to test images and flip the prediction labels with high confidence. 
Authors in \cite{w2} show that CNNs are highly vulnerable to various types of adversarial attacks. It further highlighted that achieving robustness in CNNs, using methods like adversarial training and dual batch normalization (DBN) requires precise knowledge and careful tuning to perform effectively. This work showed that vision transformers (ViTs) perform comparably to CNNs under baseline conditions but it has notable robustness against adversarial attacks even without adversarial pretraining or modifications to the architecture. This indicated the potential of vision transformers in patch-level computational pathology compared to traditional CNNs. The robustness of ViTs against input perturbation was studied in \cite{w10}. The result illustrated that ViTs remained robust even when any single layer was removed during training when there was sufficient data available. In \cite{w3}, authors evaluated the performance of the DL models by adding nine types of common corrupted pathology images into the validation set. This research introduced two classifications and one ranking metric to assess the robustness of popular CNN architectures. 

Our proposed work is fundamentally different from very few efforts for robustness in digital pathology for the following two reasons: (1) none of the existing papers have addressed robustifying against natural disturbances and purturbations on WSIs; and (2) they all focused on using patch-level images, which cannot comprehensively capture all the tumor neighborhood information in WSIs. In our previous work \cite{milan,milan2}, it was shown that graph-based algorithms lead to significant performance improvement compared to patch-based methods. Addressing these limitations, recent studies have been dedicated to employing graph-based learning methods on WSIs, demonstrating that these approaches improve the accuracy compared to the CNN-based models \cite{graph1,graph2,graph3, ref2}.\\
The use of GCNs has been very promising in tasks such as node classification, link prediction, and graph classification, but real-world graph data often contains noise, missing information, or corruptions that can negatively impact the model's performance. The work in \cite{advgraph1} is the first paper that studied the GCN's adversarial attack and robustness for node-level classification. Following this research, the field of adversarial robustness on graphs had significant growth, with numerous studies exploring various tasks, and models to enhance the robustness of GCNs against corrupted and noisy samples. Graph dropout is an adaptation of the traditional dropout technique to the graph domain. It randomly drops out nodes or edges from the graph during training. This process forces the GCN to learn more robust representations that can handle missing or noisy information effectively \cite{w1}. The study presented in \cite{w5} followed a similar approach, but it went a step further in enhancing the robustness of GCNs. The researchers enhanced the GCN's robustness by training the model to drop task-irrelevant edges through penalization of the number of edges in the sparsified graph using parameterized networks. Another technique to improve the GCN robustness is using the graph attention networks (GATs) which gives higher importance to the relevant nodes or edges during the message-passing process \cite{GAT}. By assigning attention weights to graph elements, the GCN can effectively filter out noise and focus on more informative features.

{One effective strategy for eliminating noise from graphs involves the application of graph signal processing (GSP) techniques. Research in GSP has demonstrated that incorporating a low-pass filter and implementing early stopping can be beneficial in preventing overfitting, consequently enabling the filtration of noise from graph-structured data, as detailed in \cite{denoising1}.} The study presented in \cite{w6} used GSP techniques and showed that using low-pass filters on feature vectors improves network stability. The paper in \cite{w7} used principle component analysis (PCA) as an aggregator. By utilizing PCA, the method aimed to compress neighboring node features, thereby enhancing the model's denoising capability. 

In this paper, we robustify the performance of the algorithm through a denoiser.
\begin{figure*}
    \centering
    \includegraphics[width=\textwidth]{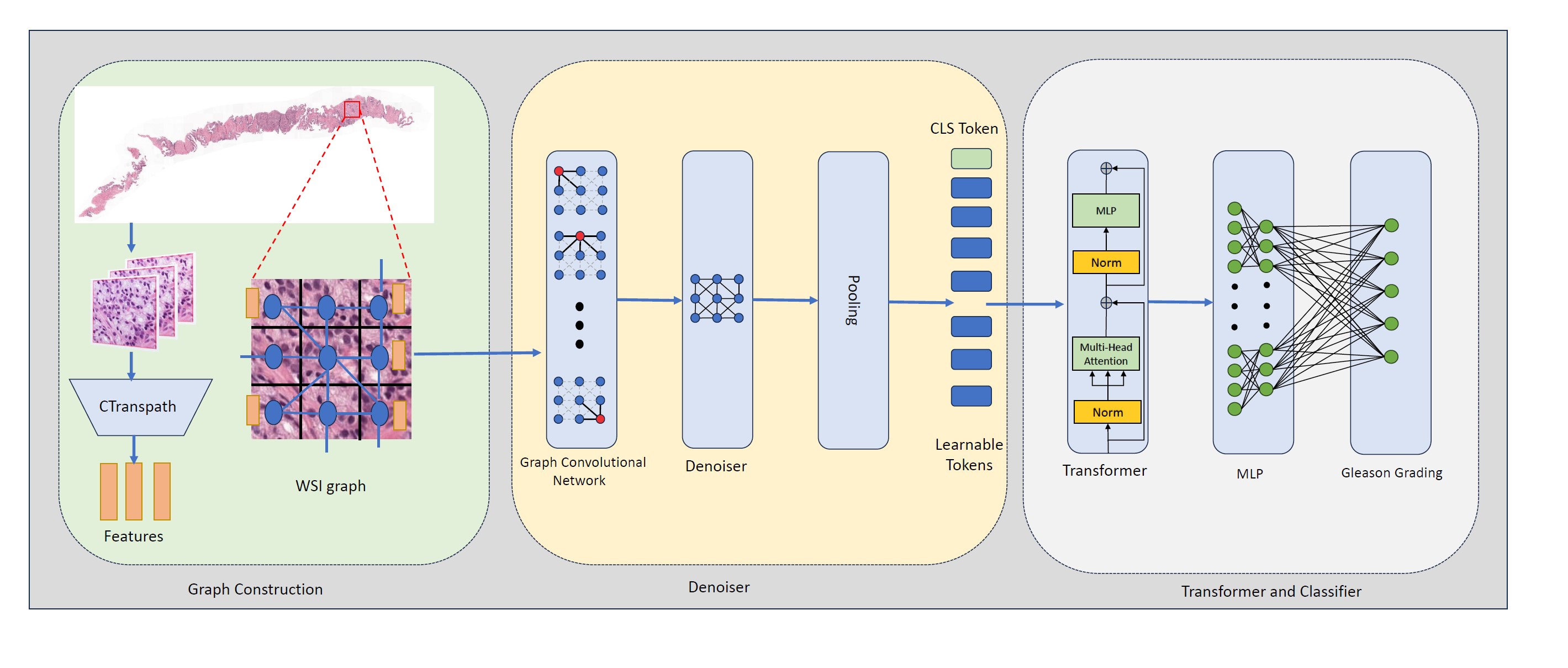}
    \caption{The overall network architecture. First, we perform data pre-processing, followed by the utilization of GCN to extract features from the graph that represents WSIs. Subsequently, these features undergo a graph denoising process and pooling layer. Finally, the denoised graph is fed into a transformer model to classify grades of prostate cancer. }
    \label{fig:model}
\end{figure*}
Perturbations on WSIs can simulate various scenarios that may occur in real-world clinical practice, such as variations in image quality, color, staining artifacts, tissue preparation inconsistencies, or differences in scanner technologies. These perturbations affect certain aspects of the image, such as altering the color balance, adding noise, simulating staining inconsistencies, or slide digitization artifacts. These perturbations can be applied globally to the entire image or localized to specific regions of interest, depending on the scenario.

In this paper, we propose a novel GCN-based architecture that is robust in handling various real-world clinical corruptions on WSIs. We initially generate real-world purturbations to the WSIs.  Through a comprehensive evaluation, we compare the performance of our proposed method against state-of-the-art GCN models, showcasing the superior robustness of our approach in managing various challenges encountered in clinical settings. The contribution of this work can be summarized as:
\begin{enumerate}
\item Evaluating the vulnerability of advanced GCN networks on managing natural noise and corruptions present in WSIs; and
\item Introducing a novel graph-based architecture designed to improve robustness in handling natural and inevitable noise and corruptions in WSIs.
\end{enumerate}

\section{Method}
In this section, we provide a detailed description of the methods employed in this study. Initially, we explain the GCN, transformers, and denoising techniques used, followed by a comprehensive overview of our proposed method. Fig. \ref{fig:model} shows an overview of our proposed method. In this work, we aim at addressing the challenges posed by WSIs, such as preserving contextual information across different image regions and handling gigapixel-sized images. To achieve this, we employed GCNs to extract features from the WSIs. Additionally, a denoising block was incorporated to mitigate the impact of WSIs artifacts like motion or blurriness.

By leveraging the capabilities of transformers, we enhanced our proposed method to accurately classify different grades of prostate cancer. The combined use of GCNs, denoising, and transformers allowed us to tackle the complexities inherent in WSIs, leading to improved performance and robustness in prostate cancer grading.

\subsection{Graph-Based Learning}

In this work, WSIs are represented as a graph. In Fig. ~\ref{fig:model} the construction of the graph from WSI is illustrated. First WSI is broken into non-overlapping patches and features is extracted for each patch. The extraction of features is performed using pretrained CTranspath~\cite{wang2022}. Each patch acts as a node and edges capture the spatial relationships or contextual dependencies between them. The edges between different nodes is connected using k-nearest neighbors (k-NN) using the coordinates of each patch in the WSI. Based on the edge connection between different nodes, the adjacency matrix for the graph is constructed. Once WSIs are represented as graphs, GCN can be employed for the learning task.

 Graph data structure consists of set of nodes $V$, adjacency matrix {$A \in \mathbb{R}^{N \times N}$} and features {$\boldsymbol{x}\in \mathbb{R}^{N}$} defined as $G(V,A,\boldsymbol{x})$. In GCN~\cite{kipf2017semisupervised} learning is done by passing messages between neighboring nodes. The message passing in each layer is given by following
\begin{equation}
    H_{l+1} = \sigma(\hat{A}H_lW_l) 
\end{equation}
where ${H_l} \in \mathcal{R}^{|V|\times d}$ is the input at the layer $l$ of GCN with each node having $d$ features, $W_l \in \mathcal{R}^{|V|\times d} $ is trainable weight matrix at layer $l$ and $\sigma$ is the non-linear activation function and $\hat{A}:=\tilde{D}^{-\frac{1}{2}}\tilde{A}\tilde{D}^{-\frac{1}{2}}$ is the symmetric normalized adjacency matrix. Here $\tilde{A} := A + I$ is adjacency matrix with added self-connection and $\tilde{D}$ is the diagonal matrix of $\tilde{A}$ with $\tilde{D}_{ii} :=  \sum_j \tilde{A}_{ij}$. 

\subsection{Transformer}

Motivated by the architecture proposed in~\cite{vaswani2023attention, zheng2022graphtransformer}, the graph transformer used in this work is presented in this section. The inputs for the transformer are represented by $H^{pool} \in \mathcal{R}^{N \times d}$ with $N$ nodes and $d$ features in each node after the pooling layer of the graph. Each node with its feature is an input token to the transformer along with a class token (CLS). Then, the transformer output is passed through a multi-layer perceptron (MLP) for classification. For each instance of WSI the predicted output $\hat{y}$ from the transformer is as follow:
\begin{equation} 
    \hat{y} = MLP(Transformer([CLS;H^{pool}])
\end{equation}
The transformer consists of multiple layers and each layer in the transformer is given by,
\begin{equation}
    t_i^{'} = MHA(LN(t_{i-1}))+ t_{i-1}
\end{equation}
\begin{equation}
    t_i = MLP(LN(t^{'}_i))+t_i^{'}
\end{equation} \\
where $MHA$ is a multi-headed self-attention, $LN$ refers to the layer normalization, and $t_i$ is the $i^{th}$ layer of the transformer. The intial layer for the transformer is $t_0 = [CLS;H^{pool}]$.
The transformer makes use of a MHA mechanism in which multiple number of self-attentions are concatenated. The equation for the MHA with $h$ heads is given by\\
\begin{equation}
    MHA = Concat[A_1,A_2,\cdots,A_h]W
\end{equation}
Here, $W$ is a trainable parameter weight matrix, $A_i$ is the $i^{th}$ self-attention. self-attention uses key(K), query(Q), value(V) mechanism which is given as follows\\
\begin{equation}
    A_i(Q,K,V) = softmax \left(\frac{QK^T}{\sqrt{d_k}}\right)V
\end{equation}\\
where $d_k = d/h$. With $W_Q,W_V,W_K \in \mathcal{R}^{d \times d_k}$ as learnable weight parameters, $Q,K,V$ used in attention follow\\
\begin{equation}
    \begin{aligned}
      Q = H^{pool}W_Q  \\
      K = H^{pool}W_K \\
      V = {H^{pool}{W_V}}.
    \end{aligned}
\end{equation}

\subsection{Denoising}
\label{Denoising}
{Supervised DL models are trained on labeled examples and aim to generalize well to unseen data. However, noise/ perturbation in the input data can hinder this generalization ability, causing the model to learn spurious or irrelevant patterns and significantly degrade the performance.

{In GCNs, denoising is essential for improving robustness due to the complex and noisy nature of graph-structured data. Denoising in GCNs helps to enhance the quality of graph-structured data and is more challenging since it can be done by removing noise or irrelevant edges, nodes, or features, uncovering the underlying structure and relationships in the graph data. This leads to more interpretable representations, enabling better understanding and trust in the learned GCN models. Implementing denoising techniques allows GCNs to identify and mitigate the impact of noise in the graph data, contributing to the creation of more accurate and reliable predictive models.}

In this work, the goal is to separate the noise {$\boldsymbol{n}\in \mathbb{R}^{N}$} from the input signal {$\boldsymbol{x}\in \mathbb{R}^{N}$} and estimate the denoised signal {$\boldsymbol{x_d}\in \mathbb{R}^{N}$} as accurately as possible. 
\begin{equation}
\boldsymbol{x} = \boldsymbol{x_d} + \boldsymbol{n}
\end{equation}}
{The problem then amounts to finding the minimum GCN wights $\boldsymbol{\boldsymbol{\theta}}$ by minimizing the following loss function: 
\begin{equation}
\label{eq9}
l(\boldsymbol{x},\boldsymbol{\boldsymbol{\theta}}) = \frac{1}{2} || \boldsymbol{x} - f_{\boldsymbol{\theta}}(\boldsymbol{z}|G)||^2_2
\end{equation}}

{GCN can be represented through a parametric non-linear function denoted as $f_{\boldsymbol{\theta}}(\boldsymbol{z}|G)$, where $G$ refers to the graph, $\boldsymbol{\theta}$ represents the network parameter, and $\boldsymbol{z}$ denotes the initial random value obtained from a zero-mean Gaussian distribution. \\
{ Leveraging the principles of graph signal processing, along with the framework of UNNPs and building upon prior work such as \cite{denoising1}, our proposed approach aims at denoising graph inputs to extract the true signal while separating it from the noise.

This method is based on the the fact that the overparameterized networks have the capacity to fit any signal, including noise and {{ alongside early stopping which aids in fitting to signals more rapidly compared to noise. Using the above mentioned method, we could extract the denoised graph ($\boldsymbol{{x}'_d}$) from the input.}}
The minimization of \eqref{eq9} can be addressed using the stochastic gradient descent (SGD) technique combined with early stopping. The values within the parameters $\boldsymbol{\theta}$ and $\boldsymbol{z}$ are initialized randomly (from i.i.d zero-mean Gaussian distribution). The weights learned after several denoising iterations on $x$ are represented as ${\boldsymbol{\theta}}'$. This strategy allows us to generate the denoised signal $\boldsymbol{{x}'_d}$ as the output corresponding to these specific weights. This approach differs from the traditional method where $\boldsymbol{\theta}$ is initially learned by fitting to the training set.\\
\begin{equation}
\boldsymbol x'_d = f_{{\boldsymbol\theta'}_{(\boldsymbol x)}}(\boldsymbol Z|G)
\end{equation}
{ This suggests that for every pair of a noisy signal $\boldsymbol{x}$ and its corresponding denoised signal $\boldsymbol{{x}'_d}$, there exists a unique set of weights associated with them, enabling us to extract the optimal denoised output for the input graph.}

\begin{figure}[ht]
\includegraphics[width=\linewidth]{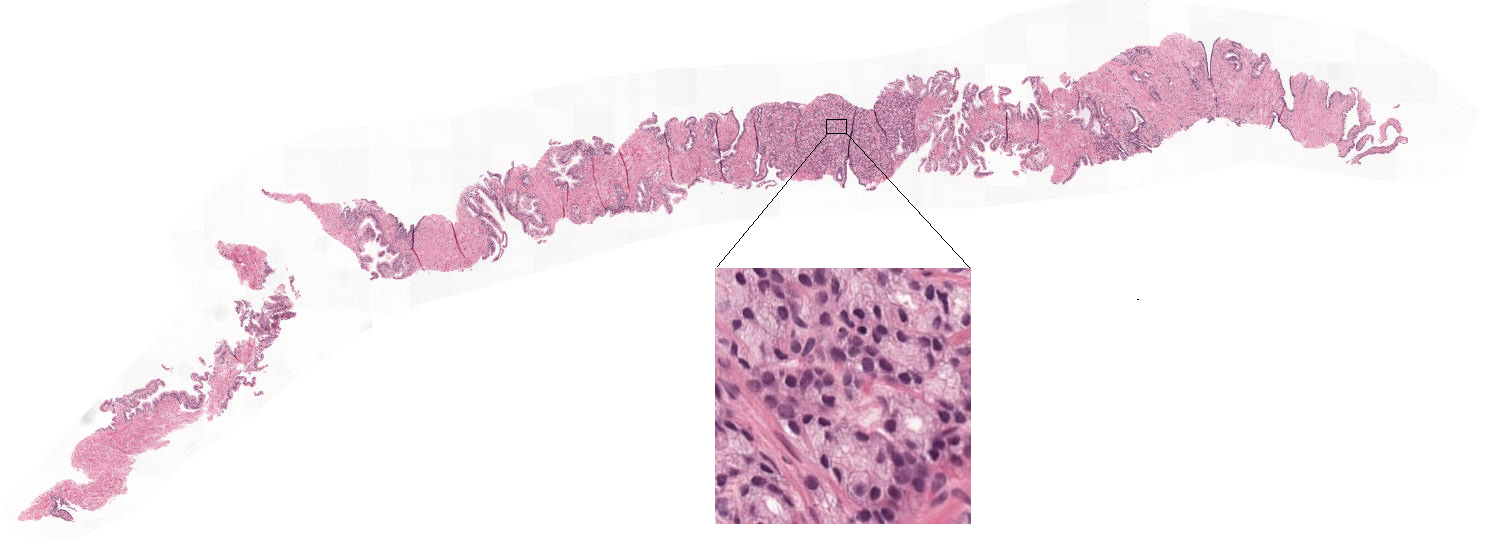}
\centering
\caption{WSI without any perturbation.}
\label{fig:original}
\end{figure}
\begin{figure*}[htp]
    \centering
    \begin{tabular}{p{0.3\textwidth} p{0.3\textwidth} p{0.3\textwidth}}
    {\includegraphics[width=0.3\textwidth]{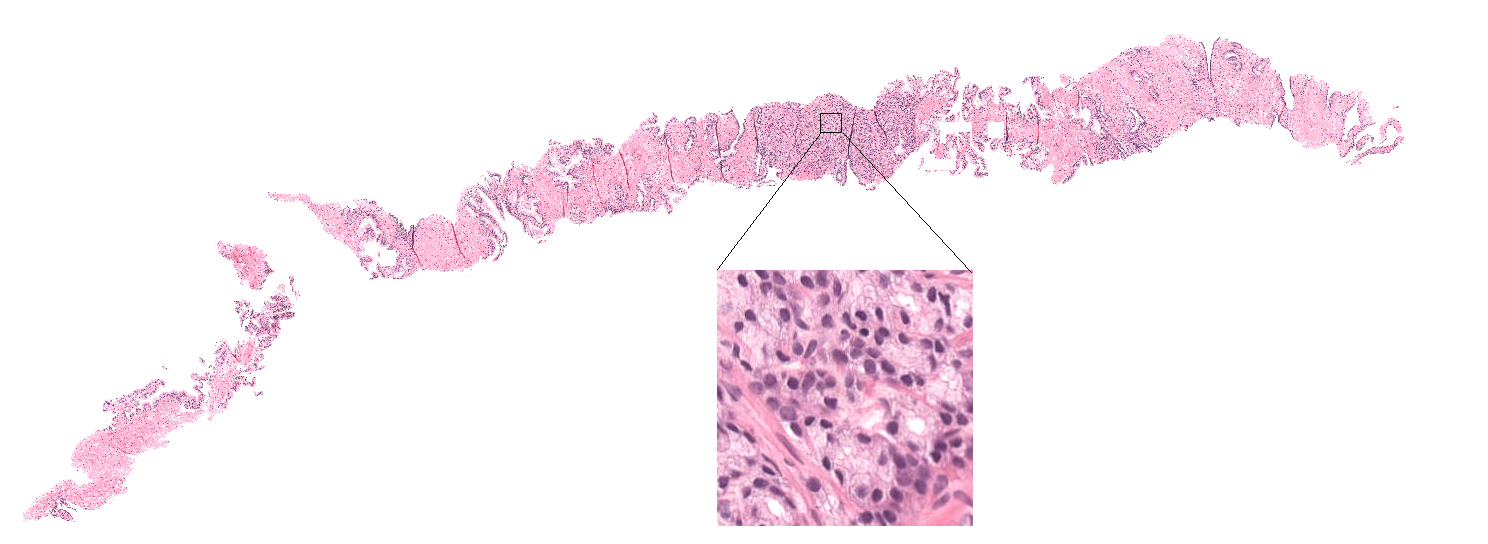}}&{\includegraphics[width=0.3\textwidth]{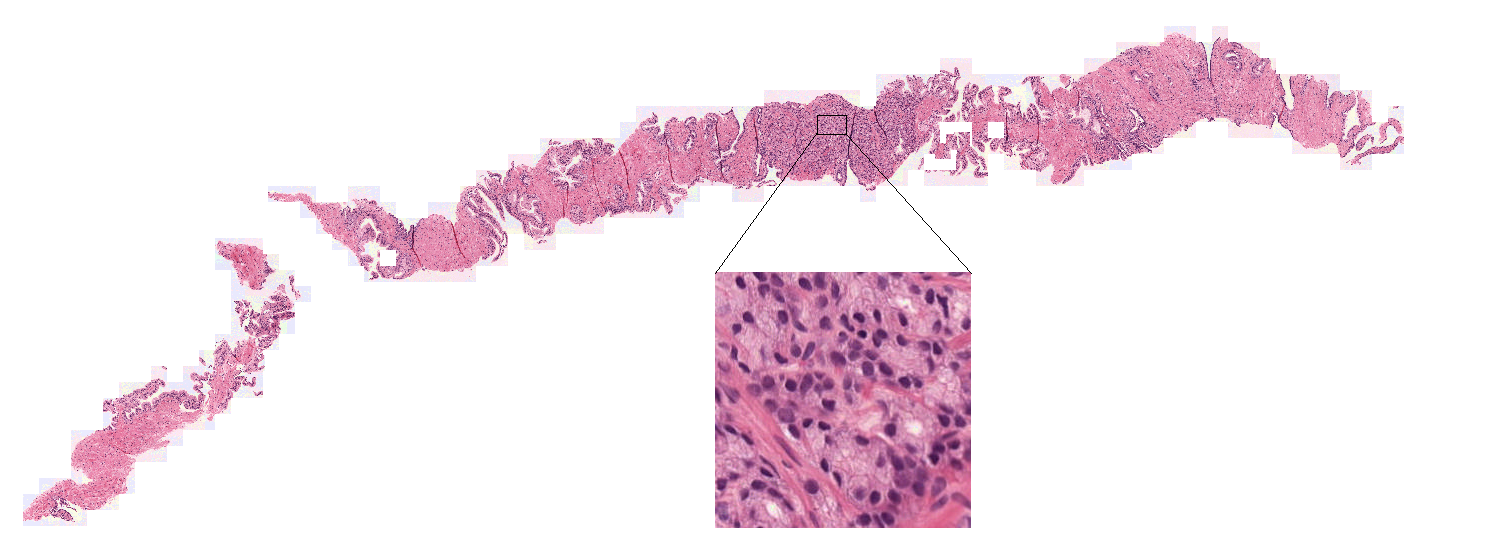}}&{\includegraphics[width=0.3\textwidth]{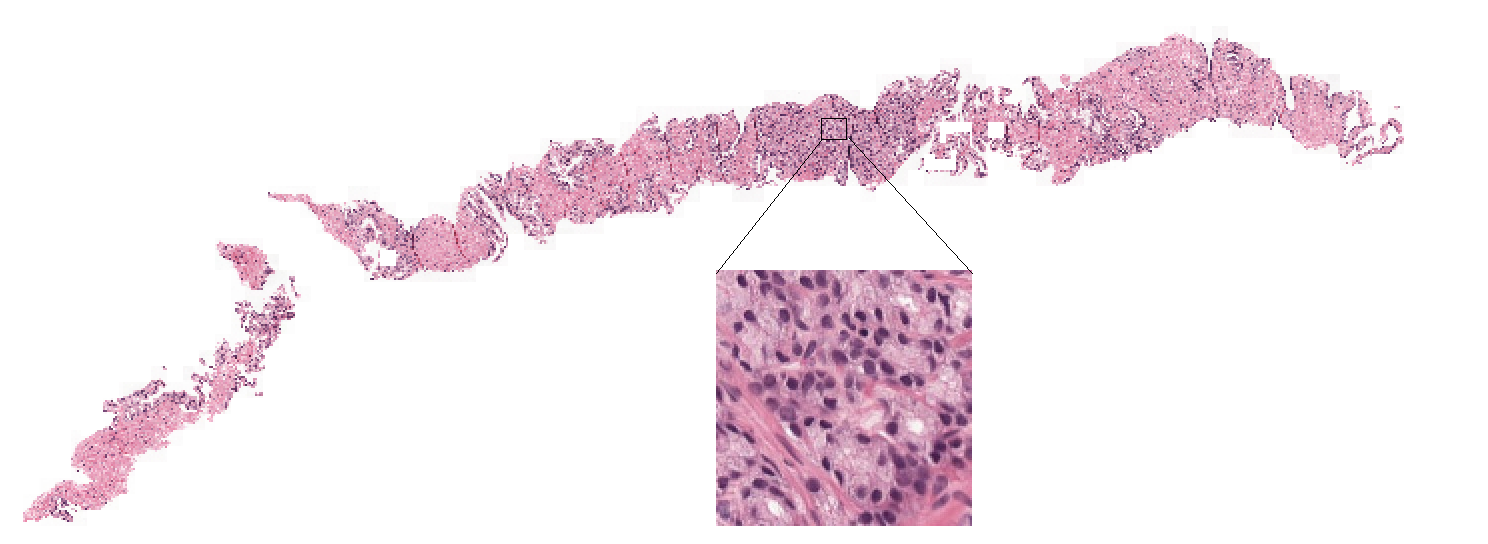}}\\
    \centering(i) Bright  & \centering(ii) Saturate & \centering(iii) Pixelate \\
     \cr
    {\includegraphics[width=0.3\textwidth]{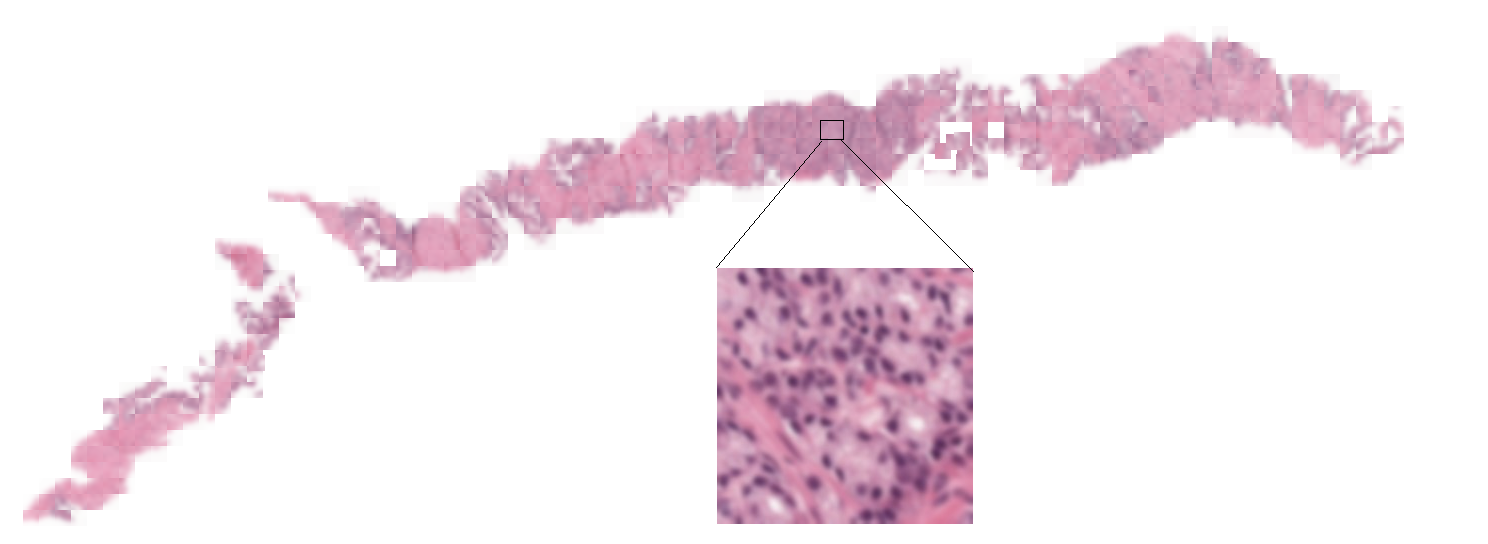}}&{\includegraphics[width=0.3\textwidth]{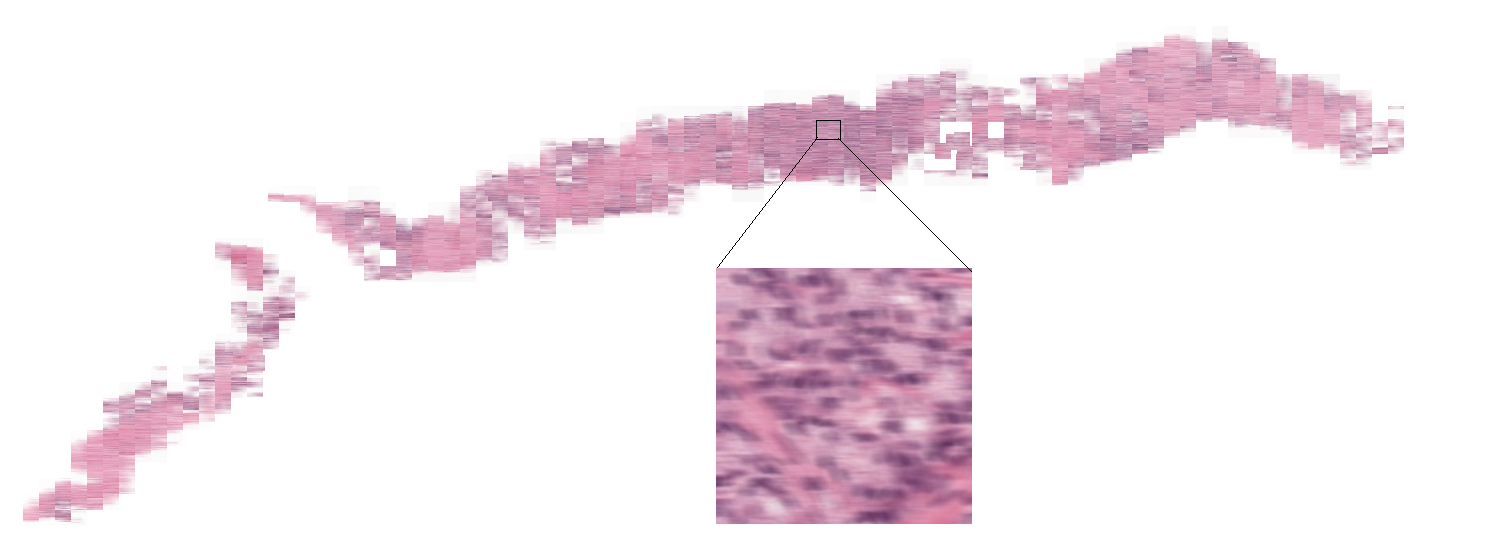}}&{\includegraphics[width=0.3\textwidth]{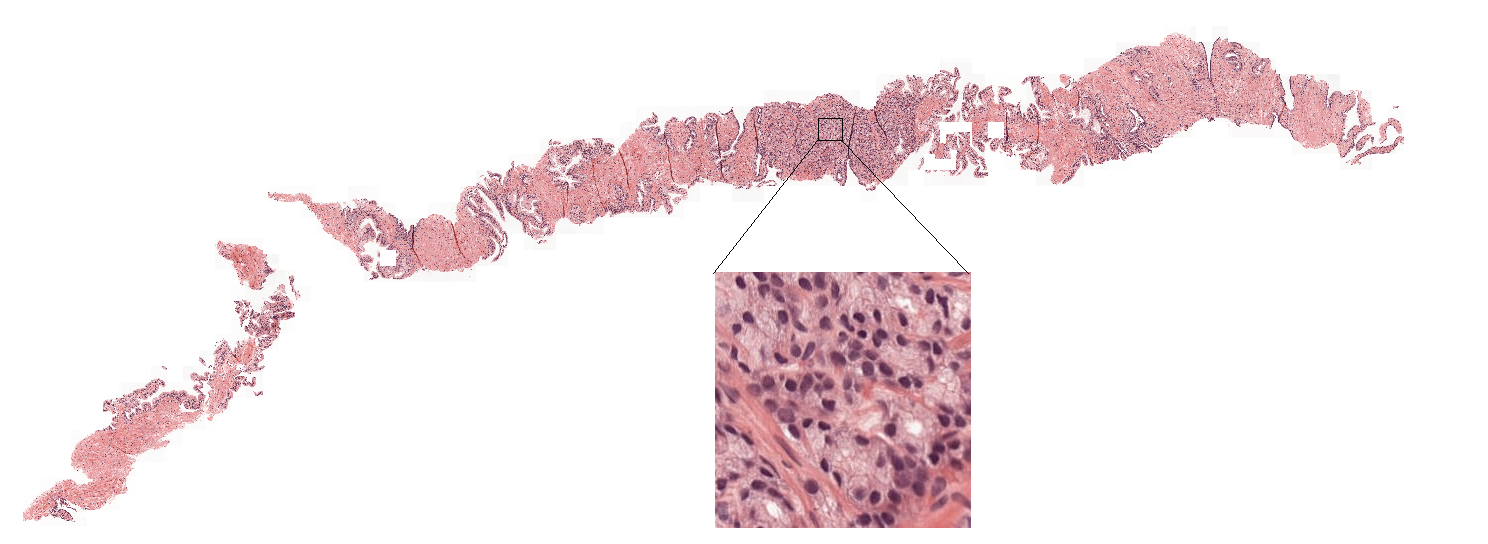}}\\
    \centering(iv) Defocus  & \centering(v) Motion & \centering(vi) Hue \\
        
  \end{tabular}
    \medskip
    \caption{Different perturbations applied to WSI in Fig. ~\ref{fig:original}.}
    \label{fig:perturbations}
  \end{figure*}

\section{Dataset}
In this work, PANDA {(Prostate cANcer graDe Assessment)} dataset consisting of 5 grades of prostate cancer {biopsies based on the Gleason score} is looked into. { Prostate cancer is the second most prevalent cancer among males globally, leading to a staggering 350,000 deaths each year~\cite{data1}. { This dataset comprises of high-resolution WSIs of prostate tissue samples obtained from biopsies primarily utilized for researching and developing solutions in the field of prostate cancer.} Skilled pathologists meticulously assess and assign scores to these tissue samples based on the Gleason grading (GG) system, a pivotal factor in determining optimal treatment strategies for patients.} \\
The dataset is available at \cite{data1} which consists of around 10000 WSIs. The breakdown of each sample is presented in Table~\ref{tab:pca}. Among these WSIs around 25\% of the WSIs are randomly chosen and different perturbations/artifacts are introduced.\\

\begin{table}[h]
\caption{Number of samples of GG in prostate cancer dataset}
\begin{center}
\begin{tabular}{|c|c|}
\hline
\textbf{Description }& \textbf{Number of samples} \\
\hline
GG1 & 2666  \\
\hline
GG2 & 1343  \\
\hline
GG3 & 1250\\
\hline
GG4 & 1242  \\
\hline
GG5 & 1224\\
\hline
\end{tabular}
\label{tab:pca}
\end{center}
\end{table}

\section{Corruption setup}
{ WSIs are digital representations of biopsy or tissue samples that have been mounted on glass slides and subsequently stained through a chemical process to enhance the visibility of their structures. This staining process, while crucial, is sensitive to various parameters including the thickness of the specimen, concentration of the stain, ambient noise, duration of staining, and the temperature at which the process occurs. Deviations in any of these parameters can lead to alterations in the appearance of the tissue sample, resulting in what is referred to as corruption in the final WSI. Such corruption denotes a degradation in the integrity and quality of the digital images, thereby posing significant challenges to their accurate analysis and interpretation.} In fact, WSIs are prone to corruption and perturbation due to (1) the complex recording and processing procedures such as tissue processing, cutting, staining, scanning, and storage. (2) inter-class differences in pathology images are smaller and blurrier than those in natural images.\\
In addition, WSIs are commonly captured in RGB color space; however, they may need to be converted to other color spaces for specific analysis or compatibility purposes. Errors in color space conversions can lead to color shifts, and information loss, hence affecting the diagnosis and analysis. To ensure diagnostic accuracy, it is important to take into account these factors and corruptions \cite{ref1,w8,w9}. Some of the common corruptions and their causes are summarized as follows:
\begin{itemize}
  \item \textbf{Brightness}: Changes in brightness can significantly impact the appearance of WSIs. Such modifications may result in mis-interpretation of color-coded information or pose challenges in distinguishing various tissue structures and cellular components accurately.\\
  \item \textbf{Saturate}: Saturation refers to various saturation intensities across different regions of the slide. This type of corruption can influence the color representation of tissues, and compromise precise diagnosis and analysis in digital pathology.\\
  \item \textbf{Pixilation}: Pixilation occurs when the resolution of the image is reduced, resulting in the loss of fine details and sharpness. This can make it challenging to discern intricate features within the image, affecting the accuracy of diagnostic tasks.\\
  \item \textbf{Defocus}: Defocus blur happens when the image is captured out of focus, leading to a lack of sharpness and clarity. As a result, important structures in the image may become unclear, making it difficult to accurately interpret the WSI. Defocus blur occurs when there is an uneven tissue thickness, or due to lens aberrations. \\
  \item \textbf{Motion}: Motion blur happens due to slide movement, scanner instability, and scanning speed during the image-capturing process, causing smudging or blurring of certain areas. This can distort critical information and hinder the ability to identify specific regions of interest in the WSI.\\
  \item \textbf{Hue}: Hue happens when the color hue is variant across different regions of the image. Hue corruption can affect the color representation of tissues and structures, potentially leading to misinterpretation and diagnostic challenges in digital pathology. Slide quality, staining variability, and scanner setting are the most common reasons for this corruption. \\
  \item \textbf{Mark}: Mark corruption refers to the presence of unwanted marks or annotations on the image that happens due to quality control or annotation verification.  Mark corruption may hide important details and structures, leading to diagnostic challenges in digital pathology. \\
\end{itemize}  
Fig. \ref{fig:perturbations} illustrates the effect of various corruptions on a sample WSI. To enhance the clarity of the illustration, we have zoomed in and shown a small patch.}\\

\begin{table*}[ht!]
\centering
\caption{Comparison of Method Performance}
\label{tab:results}
\begin{tabular}{|l|c|c|c|c|c|c|}
\hline
Method & \multicolumn{2}{c|}{Dataset} & \multicolumn{2}{c|}{25\% Noisy images} & \multicolumn{2}{c|}{Noisy Images + Denoiser} \\
\cline{2-7}
 & Accuracy & Kappa Score & Accuracy & Kappa Score & Accuracy & Kappa Score \\
\hline
Proposed Method &\textbf{0.8134} &\textbf{0.92} & \textbf{0.7563} &\textbf{0.8683} & \textbf{0.7764} & \textbf{0.8822} \\
\hline
GCN (Defferrard et al.) &0.7688&	0.887&	0.7236&	0.8457&	0.7475&	0.8662 \\
\hline
GAT (Velickovic et al.) &0.7758&	0.8909&	0.7462&	0.8561&	0.7644&	0.8648\\
\hline
\end{tabular}
\end{table*}

\section{Implementation Details}
Before implementing the proposed method, WSIs have to be prepared for processing in graph-based algorithm. Each WSI was broken into non-overlapping patches of $256\times256$ at $16 \times$ resolution. The features for these patches were extracted using CTranspath~\cite{wang2022}. The number of nodes in graph were based on the number of patches in each WSI.  

\begin{table*}[h!]
\centering
\caption{Effect of different perturbations in accuracy with and without denoiser}
\label{tab:per}
\begin{tabular}{|c|c|cc|cc|}
\hline
\multirow{2}{*}{Method}          & \multirow{2}{*}{Perturbation} & \multicolumn{2}{c|}{10\%}                    & \multicolumn{2}{c|}{50\%}                    \\ \cline{3-6} 
                                 &                               & \multicolumn{1}{c|}{W/o Denoiser} & Denoiser & \multicolumn{1}{c|}{W/o Denoiser} & Denoiser \\ \hline
\multirow{6}{*}{Proposed Method} & Bright                        & \multicolumn{1}{c|}{0.7714}       & 0.7946   & \multicolumn{1}{c|}{0.7550}       & 0.7776   \\
                                 & Hue                           & \multicolumn{1}{c|}{0.7651}       & 0.7921   & \multicolumn{1}{c|}{0.7531}       & 0.7845   \\
                                 & Motion                        & \multicolumn{1}{c|}{0.7343}       & 0.7776   & \multicolumn{1}{c|}{0.7180}       & 0.7462   \\
                                 & Saturate                      & \multicolumn{1}{c|}{0.7607}       & 0.7820   & \multicolumn{1}{c|}{0.7588}       & 0.7764   \\
                                 & Pixelate                      & \multicolumn{1}{c|}{0.7626}       & 0.7858   & \multicolumn{1}{c|}{0.7531}       & 0.7739   \\
                                 & Defocous                      & \multicolumn{1}{c|}{0.7613}       & 0.7657   & \multicolumn{1}{c|}{0.7531}       & 0.7651   \\ \hline
\multirow{6}{*}{GCN}             & Bright                        & \multicolumn{1}{c|}{0.7764}       & 0.7783   & \multicolumn{1}{c|}{0.7343}       & 0.7481   \\
                                 & Hue                           & \multicolumn{1}{c|}{0.7569}       & 0.7374   & \multicolumn{1}{c|}{0.7525}       & 0.7506   \\
                                 & Motion                        & \multicolumn{1}{c|}{0.733}        & 0.7318   & \multicolumn{1}{c|}{0.7161}       & 0.7198   \\
                                 & Saturate                      & \multicolumn{1}{c|}{0.7481}       & 0.7569   & \multicolumn{1}{c|}{0.7393}       & 0.75     \\
                                 & Pixelate                      & \multicolumn{1}{c|}{0.7494}       & 0.7626   & \multicolumn{1}{c|}{0.7487}       & 0.7569   \\
                                 & Defocous                      & \multicolumn{1}{c|}{0.7406}       & 0.767    & \multicolumn{1}{c|}{0.7481}       & 0.7531   \\ \hline
\multirow{6}{*}{GAT}             & Bright                        & \multicolumn{1}{c|}{0.7657}       & 0.7726   & \multicolumn{1}{c|}{0.7406}       & 0.7531   \\
                                 & Hue                           & \multicolumn{1}{c|}{0.7644}       & 0.7707   & \multicolumn{1}{c|}{0.7481}       & 0.7594   \\
                                 & Motion                        & \multicolumn{1}{c|}{0.7368}       & 0.7406   & \multicolumn{1}{c|}{0.7268}       & 0.7274   \\
                                 & Saturate                      & \multicolumn{1}{c|}{0.7538}       & 0.7582   & \multicolumn{1}{c|}{0.7406}       & 0.7607   \\
                                 & Pixelate                      & \multicolumn{1}{c|}{0.7544}       & 0.7651   & \multicolumn{1}{c|}{0.7494}       & 0.7607   \\
                                 & Defocous                      & \multicolumn{1}{c|}{0.7531}       & 0.767    & \multicolumn{1}{c|}{0.75}         & 0.7588   \\ \hline
\end{tabular}
\end{table*}
{ For generating the corrupted samples, we used different image processing methods and filters and using opencv library. In this step, we used different filters and kernels in order to generate motion and defocus corrupted images. For brightness, saturation and hue corruption, converting scales have been used.   }
As regards the implementation of the denoiser, our denoiser is inspired by the idea presented in \cite{denoising1,denoising2}. We employ the ReLU operator for nonlinear transformation and graph filter and graph-convolutional generator for the linear transformation. { Furthermore, alongside the denoiser, the process of graph pooling is executed to reduce the number of nodes to 100, and the result data is forwarded to a graph transformer for the purpose of classification. 

After the features were extracted the model was trained with original dataset for 60 epochs. The Adam optimizer and cross-entropy loss function were used to train the model. The learning rate was set to be $1e-3$ with weight decay of $5e-5$. Then, using these same parameters the model was trained for the noisy images with and without the denoiser. For the evaluation of the model, we used both kappa score and accuracy { which are two commonly used metrics in the field of machine learning and statistics for assessing the performance of classification models.}. Pytorch and Pytorch\_geometric was deployed as deep learning framework and the model was trained on NVIDIA Tesla V100 GPU.  \\

\section{Results}
 We evaluate the performance of our proposed model by comparing it with state-of-the-art graph-based architectures such as GCN and GAT.  Since there are no papers robustifying graph-based learning for WSIs, we focus on most popular graph-based methods and evaluate their behaviour in response to artifacts with and without our proposed denoising approach. Initially, we evaluate the performance of the model without any perturbations. The accuracy of the model without any perturbation is 81.34\% and the Kappa score is 0.92. Then, 25\% of the dataset was perturbed with different noises. The WSIs were chosen randomly and one of the perturbation was randomly applied. Then the model was evaluated using the data including noisy images. Accuracy with this noisy dataset dropped to 75.63\% and kappa score dropped to 0.8683. Then the same noisy dataset was tested with the proposed denoiser. The use of denoiser improved the accuracy to 77.64\% and kappa score improved to 0.8822. These results are summarised in Table~\ref{tab:results}. In Table~\ref{tab:results}, the comparison of proposed method with GAT-based~\cite{GAT} and a simple GCN-based model~\cite{gcn} is also presented. With and without artifacts, it can be seen that the proposed model outperforms other existing approaches.

The next step in evaluation of the model is to observe the effect of each of the perturbations to the dataset, separately. In Table~\ref{tab:results}, the dataset consists of images with all different artifcats. { We randomly perturbed 25\% of the data. Then, these noisy/perturbed datasets were evaluated based on the model with and without the presence of the denoiser. }For the case with all the perturbation the accuracy without denoiser dropped to below 77.14\% with motion perturbation having maximum drop to 73.43\% in the case with 10\% noisy dataset. Denoiser increased the accuracy to 79.46\% for the brightness perturbation and 77.76\% for the motion perturbation.  In Fig. \ref{fig:compare_3}, a comparative analysis of three methods (the proposed technique, GAT, and GCN) is presented. This comparison is carried out over varying levels of perturbation percentages. A clear observation from is that the proposed method consistently demonstrates superior performance. Notably, this superiority is evident not only in scenarios without any perturbation but also when the data experiences various degrees of perturbation. In contrast, both GAT and GCN lag behind in performance across these perturbation levels, as illustrated in Fig. \ref{fig:compare_3}. }

In the dataset with 50\% perturbed WSIs, motion perturbation had the most drop as 71.8\%. Other perturbations had the accuracy dropped to around 75\%. In the dataset with 50\%  purturbations, the denoiser improved the accuracy by around 2\% for each perturbation.  It was seen that motion perturbation had the most effect in the overall performance of the model. Motion perturbation occur due to moving of the slide when placed in the slide which causes blurriness and loss of information. This results a drop in accuracy by more than 8\% in perturbed dataset compared to original datset. The graph-based denoiser increased the performance of the model by around 4\% in both cases. 

{The detailed accuracy results for the proposed method, as well as the GCN and GAT methods, under various perturbations, with and without a denoiser, are presented in Table~\ref{tab:per}. As shown in Table~\ref{tab:per}, the novel proposed method consistently outperforms both GCN and GAT across all six distinct perturbation types, regardless of whether the perturbation level is set to 10\% or 50\%. 

Furthermore, results clearly demonstrate the impact of the denoiser on enhancing the robustness of the proposed algorithm when exposed to with six distinct artifacts. This clarifies the denoiser's important role in improving the method's robustness and adaptability under various challenging conditions. 
\begin{figure}[ht!]
\includegraphics[width=\linewidth]{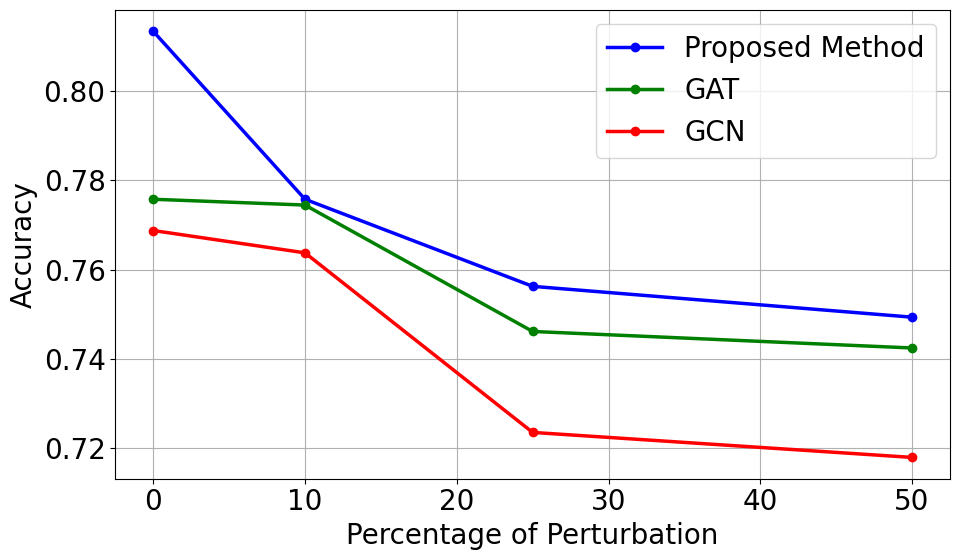}
\centering
\caption{ Model accuracy for different percentages of perturbation.}
\label{fig:compare_3}
\end{figure}

\section{Conclusion}
 The integrity of WSIs is of paramount importance for accurate diagnosis and prognosis. It is further important for medical professionals and researchers to understand the possible corruption risks associated with WSIs and enhance the network's robustness in classifying these images. By understanding and addressing these challenges, one can adopt preventative strategies to guarantee the reliability of resources in digital pathology. This work fundamentally improves the efficacy and robustness of the diagnostic process. } 

One of the key challenges in WSI classification is dealing with diverse and corrupted data. Variations in  WSIs staining, tissue preparation, and imaging conditions, lead to potential inconsistencies in the appearance of cells and tissues. 
To enhance the robustness of WSIs classification, this research proposed a novel network architecture that can process both clean and corrupted WSIs. The proposed approach involves a multi-stage process. Initially, GCN has been used to extract features from the graph representing the WSI. Subsequently, the denoiser module is used to identify and handle any potential artifacts or noise present in the WSI. Lastly, the processed features are fed through a graph transformer, enabling accurate classification of various grades of prostate cancer.

Different perturbations/artifacts on WSIs  were modelled and model's robustness to corruption and variations in WSI data was significantly improved, thus enhancing the algorithm's overall reliability in clinical applications. Tests on prostate cancer dataset verified the robustness of the proposed method when exposed to different forms and levels of perturbation. 
{ The proposed approach is both novel and innovative, particularly as it is designed to manage natural noises on the WSI-level rather than being confined to the patch-based approaches. Notably, it addresses and studies the natural disturbances and perturbations commonly encountered in the field of pathology.  With a focus on WSIs, this work provides a robust and accurate analysis of the prevalent variabilities inherent in computer-aided cancer detection, marking a pioneering step in this field.


\bibliographystyle{IEEEtran}
\bibliography{tmi}

\end{document}